\documentclass[%
reprint,
superscriptaddress,
bibnotes,
amsmath,amssymb,
prb,
]{revtex4-1}
\usepackage{float}
\usepackage[final]{graphicx}
\usepackage{amssymb,amsmath}
\usepackage{color}
\usepackage{longtable}
\usepackage{graphicx}% Include figure files
\usepackage{epstopdf}
\usepackage{dcolumn}% Align table columns on decimal point
\usepackage{bm}% bold math
\usepackage{tabularx}
\usepackage[dvipsnames]{xcolor}
\usepackage{infwarerr}
\usepackage[colorinlistoftodos]{todonotes}

\newcommand{\afm}{antiferromagnetic~} 
\newcommand{\MIT}{insulator-to-metal transition} 

\begin{document}

\preprint{APS/123-QED}

\title{Melting of magnetic order in ${\mathrm{NaOsO}}_{3}$ by fs laser pulses}% Force line breaks with \\

\author{Flavio Giorgianni}%
\email{flavio.giorgianni@psi.ch}
\affiliation{Paul Scherrer Institute, 5232 Villigen PSI, Switzerland}
\author{Max Burian}%
\affiliation{Paul Scherrer Institute, 5232 Villigen PSI, Switzerland}%
\author{Namrata Gurung}%
\address{Paul Scherrer Institute, 5232 Villigen PSI, Switzerland}
\affiliation{Laboratory for Mesoscopic Systems, Department of Materials, ETH Zurich, 8093 Zurich, Switzerland}
\author{Martin Kubli}%
\affiliation{Paul Scherrer Institute, 5232 Villigen PSI, Switzerland}
\affiliation{Institute for Quantum Electronics, Physics Department, ETH Zurich, 8093 Zurich, Switzerland}
\author{Vincent Esposito}%
\affiliation{Paul Scherrer Institute, 5232 Villigen PSI, Switzerland}%
\author{Urs Staub}%
\affiliation{Paul Scherrer Institute, 5232 Villigen PSI, Switzerland}%
\author{Paul Beaud}%
\affiliation{Paul Scherrer Institute, 5232 Villigen PSI, Switzerland}%
\author{Steven Lee Johnson}
\affiliation{Paul Scherrer Institute, 5232 Villigen PSI, Switzerland}
\affiliation{Institute for Quantum Electronics, Physics Department, ETH Zurich, 8093 Zurich, Switzerland}
\author{Yoav William Windsor}%
\affiliation{Paul Scherrer Institute, 5232 Villigen PSI, Switzerland}%
\affiliation{Fritz Haber Institut od the Max Planck Society, Faradayweg 4-6, 14195 Berlin, Germany}
\author{Laurenz Rettig}%
\affiliation{Paul Scherrer Institute, 5232 Villigen PSI, Switzerland}%
\affiliation{Fritz Haber Institut od the Max Planck Society, Faradayweg 4-6, 14195 Berlin, Germany}
\author{Dmitry Ozerov}%
\affiliation{Paul Scherrer Institute, 5232 Villigen PSI, Switzerland}%
\author{Henrik Lemke}%
\affiliation{Paul Scherrer Institute, 5232 Villigen PSI, Switzerland}%
\author{Susmita Saha}%
\affiliation{Paul Scherrer Institute, 5232 Villigen PSI, Switzerland}
\affiliation{Laboratory for Mesoscopic Systems, Department of Materials, ETH Zurich, 8093 Zurich, Switzerland}
\affiliation{Department of Physics and Astronomy,
Uppsala University, 751 21 Uppsala, Sweden}
%\author{Carlo Vicario}%
%\affiliation{Paul Scherrer Institute, 5232 Villigen PSI, Switzerland}
\author{Federico Pressacco}%
\affiliation{Physics Department and Centre for Free-Electron Laser Science (CFEL), University of Hamburg, 22761 Hamburg, Germany}
\affiliation{The Hamburg Centre for Ultrafast Imaging (CUI), 22761 Hamburg, Germany}
\author{Stephen Patrick Collins}%
\affiliation{Diamond Light Source, Harwell Science and Innovation Campus, Didcot, Oxfordshire OX11 0DE, United Kingdom}
% SACLA team
\author{Tadashi Togashi}
\author{Tetsuo Katayama}
\author{Shigeki Owada}
\author{Makina Yabashi}
\affiliation{Japan Synchrotron Radiation Research Institute (JASRI), 1-1-1 Kouto, Sayo-cho, Sayo-gun, Hyogo 679-5198, Japan}
\affiliation{RIKEN SPring-8 Center, 1-1-1 Kouto, Sayo-cho, Sayo-gun, Hyogo 679-5148, Japan}
\author{Kazunari Yamaura}
\affiliation{International Center for Materials Nanoarchitectonics (WPI-MANA), National Institute for Materials Science, Namiki 1-1, Tsukuba, Ibaraki 305-0044, Japan}%
\author{Yoshikazu Tanaka}%
\affiliation{RIKEN SPring-8 Center, 1-1-1 Kouto, Sayo-cho, Sayo-gun, Hyogo 679-5148, Japan}%
\author{V. Scagnoli}
\email{valerio.scagnoli@psi.ch}
\address{Paul Scherrer Institute, 5232 Villigen PSI, Switzerland}%
\address{Laboratory for Mesoscopic Systems, Department of Materials, ETH Zurich, 8093 Zurich, Switzerland}

\date{\today}% It is always \today, today,

\begin{abstract}
NaOsO$_3$ has recently attracted significant attention for the strong coupling between its electronic band structure and magnetic ordering. Here, we used time-resolved magnetic X-ray diffraction to determine the timescale of the photoinduced \afm dynamics in NaOsO$_3$. Our measurements are consistent with a sub-100~fs melting of the \afm long-range order, that occurs significantly faster than the lattice dynamics as monitored by the transient change in intensity of selected Bragg structural reflections, which instead show a decrease of intensity on a timescale of several ps. 
\begin{description}
\item[Usage]
Secondary publications and information retrieval purposes.
\item[PACS numbers]
May be entered using the \verb+\pacs{#1}+ command.

\end{description}
\end{abstract}

\pacs{Valid PACS appear here}% PACS, the Physics and Astronomy
\maketitle

%\tableofcontents
\section{Introduction}
The discovery of sub-ps demagnetization in ferromagnets upon ultrafast laser excitation~\cite{BeaurepairePRL76} has triggered an intense wave of research focusing on understanding the fundamental mechanisms involved in the dissipation of the spin and orbital angular momenta~\cite{BattiatoPRL105,MuellerPRL111,TurgutPRB94,HofherrPRB96,RitzmannPRB101,DornesN2019}. This fundamental research interest in the ultrafast manipulation of magnetic order is complemented by its potential applications for high-speed data storage and processing technologies, as well as its relevance for faster spintronic architectures; where, for example, one major goal of research on these systems is to develop new, ultrafast methods of switching between metastable magnetic states.~\cite{StanciuPRL99_07,LambertS_14,SchubertAPL104_14,El_Hadri_2017,BeensPRB100_19,CiuciulkaitePRM4_20}

Recent research efforts within ultrafast magnetism have been partly focused on antiferromagnets~\cite{WadleyS351,BodnarNC2018,PRA9.064040,MeerNL2021,ChenNM2019} since they possess resonant frequencies in the terahertz (THz) range, which is three orders of magnitude higher than observed for ferromagnets~\textcolor{black}{(see Ref.~\onlinecite{NemecNP14} and reference therein)}. Furthermore, in these systems the angular momentum can be directly exchanged between the spin up and spin down magnetic sublattices. Having equivalent stable states with zero net angular momentum in the spin system, antiferromagnets are expected to exhibit faster dynamics, different from those observed in ferromagnets, where momentum transfer to the lattice occurs. Given the ubiquity of \afm materials, they offer a rich playground for investigating ultrafast spin dynamics.
Of particular interest is the fact that in the majority of \afm systems the magnetic ordering is intimately related to the electronic structure of the material. Therefore by manipulating the electronic structure, one also affects the \afm order parameter.
Indeed, theoretical predictions~\cite{WernerPRB86,MentinkPRL113,EcksteinSR2016} suggest that in \afm materials falling in the weak and strong electron coupling regime quenching of the \afm order parameter occurs concomitantly within the photo-excited dynamics of the electronic system, which implies a few-fs or faster time scales are possible.
\textcolor{black}{Laser-induced ultrafast reorientation or switching of the \afm order parameter has been demonstrated in several materials~\cite{KimelN429,KimelNP5,KandaNC2,SatohNP9,KimelPRL89, KampfrathNP2011,YamaguchiPRL105, BaierlNP10,NemecNP14}, including Mott insulators~\cite{KimelPRL89,KandaNC2,KampfrathNP2011}. However, there have been a limited amount of reports that discuss the ultimate time scale of the \afm quenching in strongly-correlated systems following an ultrafast photo-excitation.~\cite{KimelPRL89,CavigliaPRB88,AfanasievPRX9,Mazzonearxiv,Stoicaarxiv} In these reports, the time scales have been limited either by the experimental time resolution or found to be in the 100-400~fs range, which is significantly slower when compared with the values predicted theoretically~\cite{WernerPRB86,MentinkPRL113,EcksteinSR2016} or reported for ferromagnetic materials and multilayer films~\cite{BeaurepairePRL76,HohlfeldPRL78,SchollPRL79,Stamm2007,KirilyukRMP82,Rudolf2012}}. It is therefore interesting to ascertain on which timescales the electronic and magnetic properties can be modified, in particular in systems in which \afm~order occurs concomitantly with an abrupt change in the electronic properties of the material, as occurring at an \MIT.

In this work we report the ultrafast manipulation of the \afm order and the electronic structure in NaOsO$_3$. The photoexcitation of NaOsO$_3$ by femtosecond laser pulses with photon energy above the insulating gap drives simultaneously the dynamics of electrons and spins. To uniquely access the laser-induced ultrafast spin dynamics we use time-resolved femtosecond X-ray diffraction on a magnetic Bragg peak. Our results show that the manipulation of the \afm order parameter occurs on a sub-100 fs time scale. 

NaOsO$_3$ undergoes an \MIT~concomitant with antiferromagnetic ordering at $T_\mathrm{IM}=T_\mathrm{N}$= 410(1)~K.~\cite{ShiPRB80} In this compound, the absence of crystallographic symmetry breaking~\cite{CalderPRL108,GurungPRB18} is suggestive of a magnetically driven \MIT. However, due to the presence of energetically similar competing interactions, a consensus on the nature of the metal-insulator mechanism operating in this perovskite is absent~\cite{CalderPRL108, middey2014, GurungPRB18, KimLifshitzPRB94, vale2018evolution, vale2018crossover}. 
\textcolor{black}{In the insulating phase, with a gap at low temperature of 102(3)~meV~\cite{LoVecchio2013}}, the magnetic moment determined by neutron diffraction refinement is $1\,\mu_{B}$, and it suggests a  coexistence of localized and itinerant magnetism~\cite{CalderPRL108}. Below $T_\mathrm{N}$, magnetic moments order almost parallel to the $c$-axis in a G-type antiferromagnet with a very small ($< 0.01~\mu_\mathrm{B}$) ferromagnetic component along the $b$-axis. 
The strong enhancement of magnetic diffraction at the Os $L_3$ edge \textcolor{black}{(10.787~keV)} makes this material very appealing for time resolved X-ray diffraction measurements of the \afm dynamics. The latter can be related to the changes in the conductivity response of the material upon fs laser excitation, which are expected to drive an \MIT~in the system.

This paper is organized as follows: in Sec.~\ref{ExpDet} we describe the sample preparation and the experimental details of the pump-probe experiments. \textcolor{black}{In Sec.~\ref{ResultsX}, we present the experimental results describing the \afm dynamics in a time window of few ps subsequent to a sub 100~fs laser excitation. These results are analysed with a time-dependent order parameter model that was applied  previously to manganite materials.~\cite{Beaud2014} We find that the melting of the \afm order following a fs laser excitation occurs on a timescale comparable with the experimental time resolution and faster compared to other 5$d$ oxide materials. Lattice dynamics, monitored via the transient change of the intensity of structural Bragg peaks, occur on a time scale of several ps.}

\begin{figure}[t]
\centering
\includegraphics[width=\columnwidth]{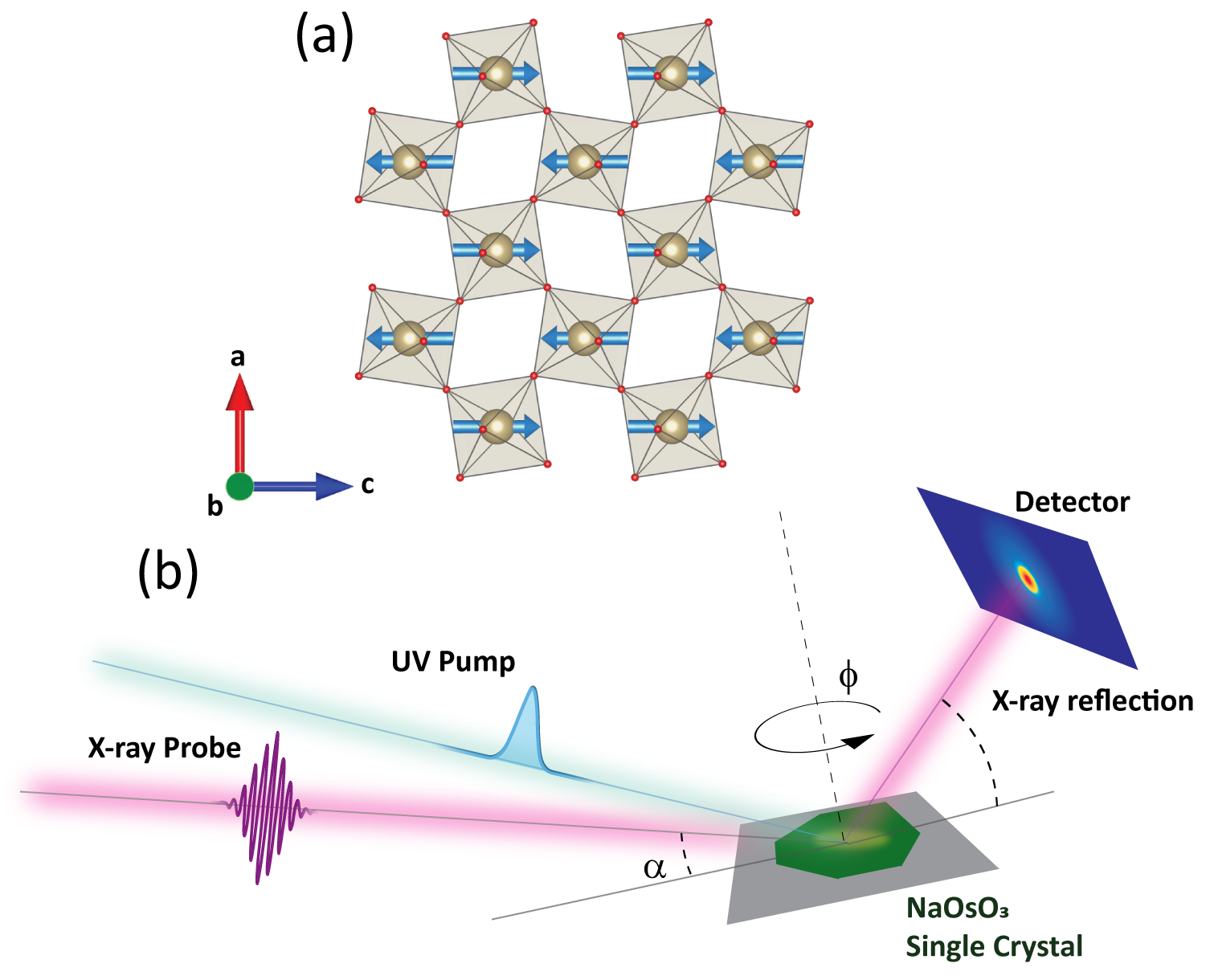}
\caption{Time-resolved resonant X-ray diffraction experiment. a) Antiferromagnetic order of NaOsO$_3$ for T$<$T$_N$ in the $a$,$c-$ plane. The Os magnetic moments (blue arrow) lying along the $c$-axis. The OsO$_6$ octahedral cages are colored in grey. b) Schematic setup of the optical-pump/X-ray-probe experiment. The experiment was performed in grazing incidence geometry $\alpha<1\deg$ and with the UV pump laser (400~nm) at an angle of 7 $\deg$ with the X-rays. \textcolor{black}{The angle $\alpha$ ($\beta$) is the angle between the incident (scattered) X-rays} and the sample surface. The angle $\phi$ rotates the sample about its surface normal.
}
\label{fig:exp_geo}
\end{figure}

\section{Experimental details} \label{ExpDet}
The resonant X-ray diffraction experiment was carried out at the EH2 endstation of the BL3 beamline at the SACLA X-ray free electron laser~\cite{IshikawaNP6}, using a Huber four-circle ($2\theta$, $\theta$, $\chi$ and $\phi$) diffractometer  \textcolor{black}{in horizontal scattering geometry equipped with a single module MultiPort Charged Coupled Device (MPCCD) detector~\cite{KameshimaRSI85}}. An illustration of the experimental setup is shown in Fig.~\ref{fig:exp_geo}. In the experiment, a NaOsO$_3$ single-crystal sample with a surface normal direction close to [1 0 0]$_{Pbnm}$ and an area of $\sim$ 100 $\times$ 100 $\mu$m$^2$  was mounted on a goniometer with a nitrogen cryostream to stabilize the temperature at approximately 293~K. A $\hat{\tau}_L=$65~fs full width at the half maximum (FWHM) optical pulse with wavelength 400~nm excited the sample with a repetition rate of 60~Hz. The laser spot size at the sample position was measured to be 300~$\mu$m $\times$  300~$\mu$m. The horizontal polarized X-ray beam operating at 30~Hz and with a pulse duration of $\hat{\tau}_{X} =$10~fs (FWHM) was focused to a spot size of 10~$\mu$m $\times$ 10~$\mu$m. During the experiment, the angle between the X-ray- and laser- beams was kept fixed at 7~degrees. An X-ray grazing incidence geometry ($\alpha=0.5^{\circ}$) was used to match the X-ray and laser penetration lengths, with the latter estimated to be 50~nm from Ref.~\onlinecite{LoVecchio2013} for a powder sample. Independent shutters for the laser and X-rays beams were used to collect data with and without laser excitation.

The X-ray beam energy was tuned within the vicinity of the Os L$_3$ edge ($2p-5d$ transition) at around 10.787~keV ($\lambda=$0.115~nm). The temporal jitter between X-ray and optical laser pulses was measured shot-by-shot using a transmission grating based timing tool, which has an accuracy $\Delta t_{TT} =$10~fs (FWHM)~\cite{KatayamaSD3}. The temporal fingerprint of each shot was then used to re-bin all data into  $\Delta t_{bin} = 50$~fs time-segments~\cite{NakajimaJSR25}. The effective time-resolution of the experiment was estimated to be $\Delta t_{\textrm{eff}}=\sqrt{ \hat{\tau}_{L}^2 + \hat{\tau}_{X}^2 + \Delta t_{TT}^2 + \Delta t_{bin}^2 } = 80(5)$~fs  (FWHM). As it is customary to model time traces using an erf($t$/$\tau$) function~\cite{Glownia:10,GierzPRL114}, which is characterized by a timescale $\tau$ which is different from the timescale $\hat{\tau}$ expressed in terms of the FWHM, the experimental time resolution $\tau_{eff}$ can also be expressed as $\tau_{\textrm{eff}}$=$\Delta t_{\textrm{eff}}/\sqrt{4\ln(2)}$= 51(4)~fs. \\
% sqrt(65^2+15^2+10^2+50^2)=83.9 fs
% sqrt(65^2+15^2+10^2+25^2)=71.9 fs
% see gaussian_erf.m

\section{Results and Modelling} \label{ResultsX} 
%\subsection{XFEL Pump probe} 
In order to monitor the time evolution of the antiferromagnetic order parameter in response to an ultrafast laser excitation, we have taken advantage of the enhancement of the X-ray magnetic cross section in the vicinity of the Os $L_3$ edge. Based on the experimental geometric constraints, we have selected the (1 $\bar{5}$ 0) magnetic reflection as the most suitable one with which to perform our pump-probe measurements.
\begin{figure}[b]
\centering
\includegraphics[width=\columnwidth]{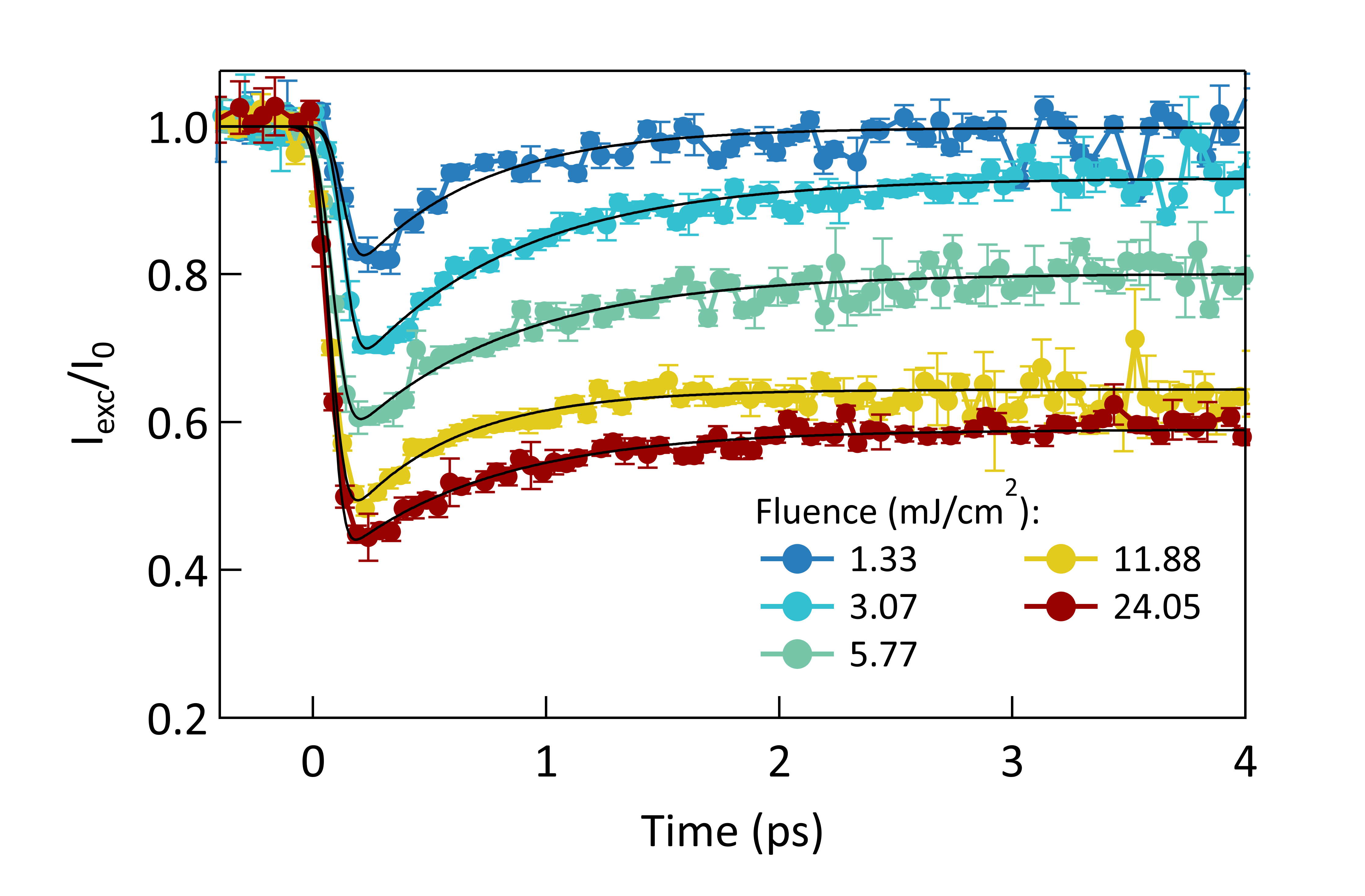}
\caption{Normalized diffraction intensity changes of the (1 $\bar{5}$ 0) magnetic reflection at the Os L$_3$ edge for several laser fluence at T=293~K. Error bars reflect the standard deviation of the data within each bin. Solid lines are fit to the data obtained with Eq.~\ref{eq:erf_rec}.}
\label{fig:AFM_fluence}
\end{figure}
The time evolution of the (1 $\bar{5}$ 0) peak maximum intensity upon optical laser excitation is illustrated in Fig.~\ref{fig:AFM_fluence} for several laser excitation fluences. In all of the time traces a sudden decrease of intensity is observed within 100~fs after the optical laser excitation. Angular $\phi$-scans (rocking curves) at fixed time delay (see Fig.~\ref{fig:phiscans}) confirm that the magnetic peak position in reciprocal space does not change within the first 10~ps after excitation. Therefore, we observe a reduction of the diffracted intensity related to the reduction of the sublattice magnetization up to 60\% on a sub ps timescale, followed by a recovery of the intensity on a several~ps timescale. The persistence of a sizable part of the magnetic diffraction intensity is likely due to the imperfect matching of the laser and the X-ray penetration depths in a crystal as observed, for example, in similar experiments~\cite{CaviezelPRB86,Windsor2020, Mazzonearxiv}. 

In order to extract an upper bound for the relevant timescales from these time traces and to readily compare such values with other experimental results~\cite{AfanasievPRX9,Mazzonearxiv,CavigliaPRB88,Stoicaarxiv}, we have fit the data using an error function, which captures the fast decay time ($\tau$), multiplied by an exponential term that captures the recovery time ($\tau_{rec}$). \textcolor{black}{This model is appropriate when the response of the material to the excitation is linear and significantly faster than the time resolution of the experiment, which in this case is determined by the pump and probe pulse duration as well as their relative timing stability.} The fitting function has the following analytical form:

\begin{equation}
f(t)= \frac{1}{2}\Big[ \text{erf} \Big(\frac{t-t_0}{\tau} \Big)+1 \Big]\cdot \Big[ A\, e^{-(t-t_0)/\tau_{rec}}+c \Big] \, ,
\label{eq:erf_rec}
\end{equation}

where $A$, and $c$ are fit parameters that represent respectively the amplitudes of the intensity reduction and the 
long-lived transient, which lasts well beyond the 4~ps time window. Here, $t_0$ represents the time where the X-rays and the optical laser impinge concomitantly on the sample. \textcolor{black}{In order to reproduce accurately the time traces, we have found that $t_0$ must be treated as a fit parameter. The shift in $t_0$ we observe as a function of the excitation laser fluence is possibly due to  an uncontrolled drift of $t_0$ during the experiment or to saturation effects. The latter would lead also to a change in $\tau$ as a function of fluence, that we do not observe outside experimental uncertainties.}

\begin{table}[t]
\caption{\label{tab:table0}%
Parameters obtained by  fitting the experimental data for the magnetic (1 $\bar{5}$ 0) reflection with Eq.~(\ref{eq:erf_rec}). Numbers within brackets represent standard deviations. $F$ stands for fluence.
}
\begin{ruledtabular}
\begin{tabular}{lccccc}
$F$~(mJ/cm$^2$) & $t_0$~(fs) & $\tau$~(fs) & $A$ & $\tau_{rec}$~(ps) & $c$         \\
1.33  & 132(8)    & 69(9) & -0.11(6)    & 0.54(5) & -0.001(3) \\
3.07  & 133(4)    & 69(7) & -0.13(3)  & 0.71(6) & -0.034(4) \\
5.77  & 85(3)    & 78(11) & -0.12(5)  & 0.71(7) & -0.099(4)   \\
11.88  & 74(3)    & 75(4) & -0.10(5) & 0.47(5) & -0.180(5)  \\
24.05  & 72(2)   & 63(5) & -0.09(5)  & 0.66(4)  & -0.207(3) 
\end{tabular}
\end{ruledtabular}
\end{table}

Taking into account the caveats above, we observe that the fit of a time trace returns a value of $\tau$ on the order of 65-75~fs for the fast decay constant (see Table~\ref{tab:table0}). Therefore, our experiments reveal the presence of a sub-100~fs drop of the magnetic peak intensity, for all the laser fluences, with an average decay time $\tau=\tau_{\text{AFM}}$ = 71 $\pm$6~fs. In addition, we observe two distinct behaviours as a function of the laser fluence. For fluences below 5~mJ/cm$^2$ the antiferromagnetic ordering recovers almost completely to the initial intensity with the first 2-3~ps, while for higher laser fluences the intensity remains suppressed over the time window of our measurements. A similar behaviour has been observed in other \textcolor{black}{transition metal based oxide} materials~\cite{CavalleriPRB87,BeaudPRL103,deJongNM12,EspositoPRB97,PorerPRB101,AfanasievPRX9} and  is often assumed to herald the occurrence of a phase transition in the probed sample volume.
In order to obtain a more quantitative description of our measurements we apply a modified version of the model presented in Refs.~\onlinecite{Beaud2014,EspositoSD2018,BurianPRR3} based on an effective time-dependent order-parameter $\eta$, which in our case represents the staggered magnetization associated with the antiferromagnetic ordering.

Within this model, there exists a threshold value $n_c$ of the absorbed local energy density per volume $n$ above which the system undergoes a phase transition where the order parameter vanishes. In our case this will correspond to the melting of the long range antiferromagnetic order when the ratio $n/n_c > 1$, i.e. when the sublattice magnetization vanishes.
The time-dependent order-parameter model can take into account the effective time resolution of the experiment $\Delta t_{\textrm{eff}}$ and the fact that the optical laser pulses are absorbed as they propagate through the sample.

Specifically, we account for the depth-dependent excitation profile by splitting the sample into N~=~400 layers of thickness of $\Delta z=$1~nm. The electronic excitation density $n_i$ of a layer at a depth $z_i$ is then $n_{0i}= n_0 \, e^{-z_i/z_L}$, where $z_L$ is the effective laser penetration depth. For a given laser fluence $F$, $n_0 = F/z_L$.
As explained in detail in Appendix A, we express the diffracted intensity $I_{exc}(t)$ as
\begin{equation}
 I_{exc}(t)/I_0 \propto  \sum_{i=0}^{N} \left|\eta(z_i,t) \right|^2,   
\end{equation}
where %$d_X$ is the X-ray electric field penetration depth and 
$\eta(z_i,t)$ is the time-dependent order parameter \textcolor{black}{normalized to unity at time before excitation and for times during and after the excitation} given by
\begin{equation}
\eta(z_i,t) = \Big(1 -  \frac{\text{min}(n(z_i,t),n_c)}{n_c} \Big)^{\gamma_0}  \, ,    
\end{equation}

where $n_c$ is a critical excitation density and $n(z_i,t)$ is the absorbed local energy density per volume, which depends on the incoming pump fluence $F$ and $\gamma_0$ is analogous to a critical exponent of the initial excitation. \textcolor{black}{If $n(z_i,t) > n_c$ the phase transition to the paramagnetic state occurs and $\eta=0$. In order to account for the recovery of the order parameter, an energy  dissipation term is introduced in the expression for $n(z_i,t)$~\cite{Beaud2014}. 
\begin{equation}
n(z_i,t) = \Big( n_0(z_i,t) - \alpha n_c \Big)e^{-\frac{t}{\tau_{rec}}} + \alpha n_c , 
\label{eq:CED}
\end{equation}}
In using this form for $n(z_i,t)$ we approximate the relaxation of the electronic energy density as a two-step process.  The time constant $\tau_\textrm{rec}$ characterizes a fast relaxation via electron-phonon interactions, which is followed by a much slower relaxation characterized by a time constant much larger than the measurement time window.  Here~\cite{Beaud2014,PorerPRB101}  
\begin{equation}
\alpha=1- \Big(1-\frac{n_0(z_i,t)}{n_c} \Big)^\frac{\gamma}{\gamma_0},
\end{equation} 
where $\gamma$ is an effective critical exponent of the quasi-thermalized system after the initial relaxation process.

The quantity $\alpha$ is a function of $\gamma_0$ and $\gamma$, which can be regarded as the critical exponents with respect to the initial excitation ($t \approx 0$) and after equilibration ( $t >> \tau_{rec}$). If $\gamma_0=\gamma$, $\alpha=n_0(z_i,t)/n_c$ and the variation in the observed intensity will be the same at short and large time after the laser excitation. If $\gamma_0>\gamma$, the change in the measured intensity will be larger at the shorter time scales, signalling the presence of a mechanism leading to a partial recovery of the order parameter with a time constant $\tau_{rec}$. For $\gamma_0>>\gamma$, one could anticipate an almost complete recovery of the order parameter, at least for the lowest excitation fluences.

We assume for the initial electron energy density
\begin{equation}
n_0(z_i,t)dz= \frac{F}{2}  \Big( 1-e^{-\frac{dz}{z_L}} \Big) e^{-\frac{z_i}{z_L}}   \Big[ 1+ \text{erf} \Big(\frac{t}{\tau_L+\hat{\tau}} \Big) \Big],
\end{equation}
which describes the energy deposited by the pump laser pulse of duration $\tau_L$ at a given depth $z_i$.
\begin{figure}[b]
    \centering
  \includegraphics[width=\columnwidth]{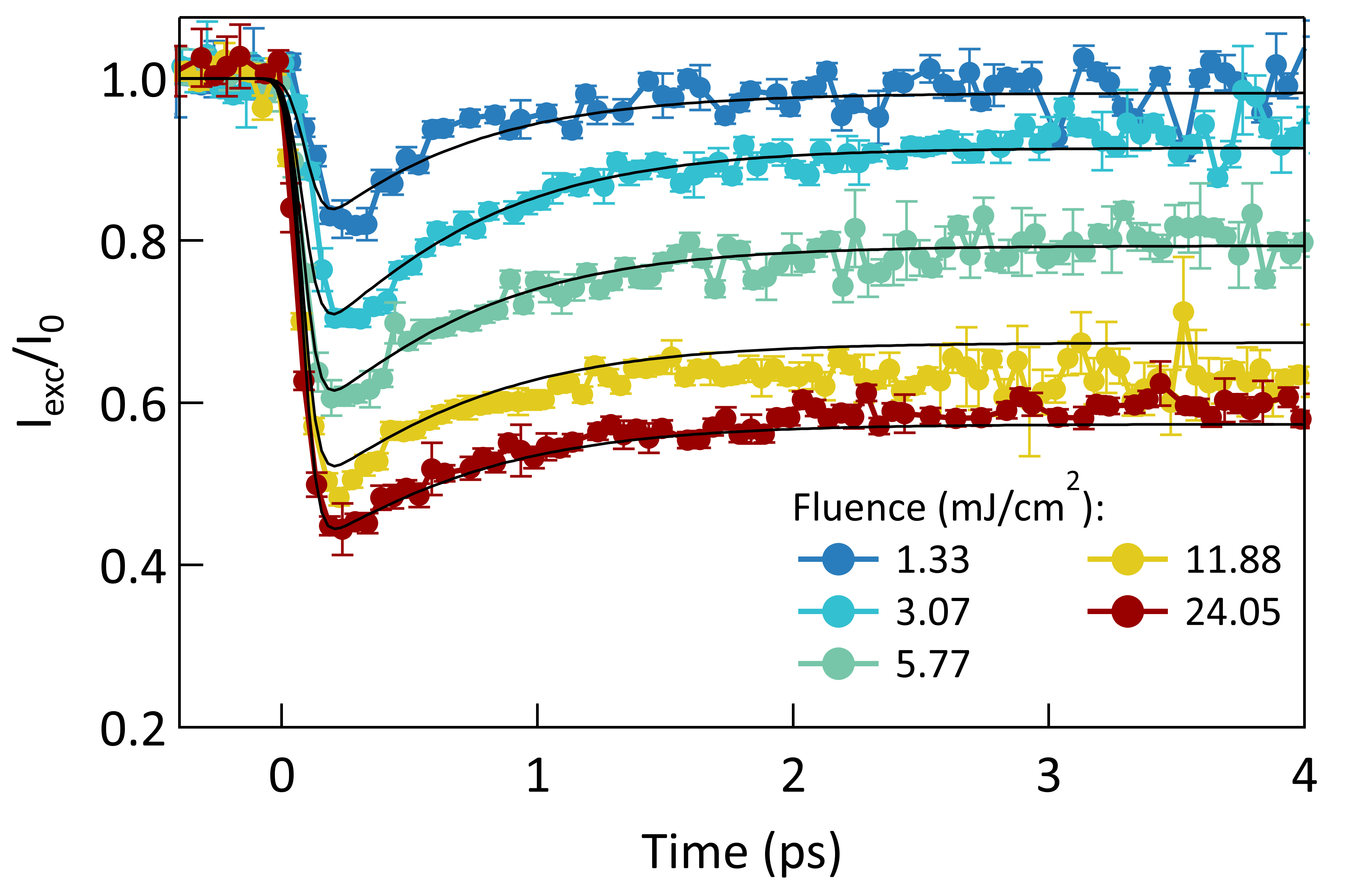}
    \caption{Fits to the time-dependent magnetic reflection data using the time dependent order parameter model described in the text for different laser pump fluences.}
    \label{fig:Maxmodel}
\end{figure}
With such a model we are able to describe reasonably well all the observed time traces, as shown in Fig.~\ref{fig:Maxmodel}. \textcolor{black}{ From the simultaneous fit of all the time traces, we find the following values (and associated uncertainties) for the fit parameters: the effective optical penetration depth $z_L= 20.9(8)$~nm, which is reasonable if we consider that the effective optical penetration depth is reduced in our geometry when compared to the 50~nm value estimated for a powder sample in normal incidence conditions~\cite{LoVecchio2013}. The critical excitation density is $n_c=813(3)$ J/cm$^3$ and the critical exponent is $\gamma_0=1.28(7)$, which is considerably higher than the values 0.5 and 0.69(3) reported for PrCaMnO$_3$~\cite{Beaud2014} and for PrCaMnO$_4$~\cite{PorerPRB101}, respectively. The estimated value for the $\gamma_0$ exponent in this experiment suggests an approximately quadratic relation between the scattering intensity and the excitation energy density. The value of $\gamma=0.08(1)$ is smaller than that  reported for PrCaMnO$_3$~\cite{Beaud2014} and PrCaMnO$_4$~\cite{PorerPRB101}, where it was found $\gamma=0.20(1)$ and $\gamma=0.29(3)$, respectively. The differences between the fitted values of both $\gamma$ and $\gamma_0$ for the present experiment and these previous studies may be partially due to the fact that the previous studies were on thin films, whereas the current study is on a bulk sample where the laser excitation depth is smaller than the x-ray penetration depth.  Since the model does not explicitly treat energy transport effects, the fitted values of the parameters may be influenced by these effects. Nevertheless, the ratio of $\gamma$/$\gamma_0$ is significantly smaller for NaOsO$_3$, making plausible to conclude that the electronic energy dissipation term is more effective in NaOsO$_3$ as it is characterized by a relaxation constant $\tau=0.48(3)$~ps, whereas in PrCaMnO$_3$ it was $\tau=0.81$~ps. Such results could possibly reflect the difference in the phase transition occurring it those two classes of materials, with the manganites undergoing a change in the structure of the material, while this should not be the case for NaOsO$_3$~\cite{CalderPRL108}. In such a scenario, one would expect a recovery requiring a lattice rearrangement to occur on a longer timescale than a reordering of spins to their \afm ground state ordering. The observed recovery time-scale $\tau_{rec}$ points to the presence of an efficient thermalization process of the transient free carrier population induced by the pump pulse, consistent with the small gap and the partially delocalized nature of the magnetism in NaOsO$_3$. }
\begin{figure}[b]
\centering
\includegraphics[width=0.9\columnwidth]{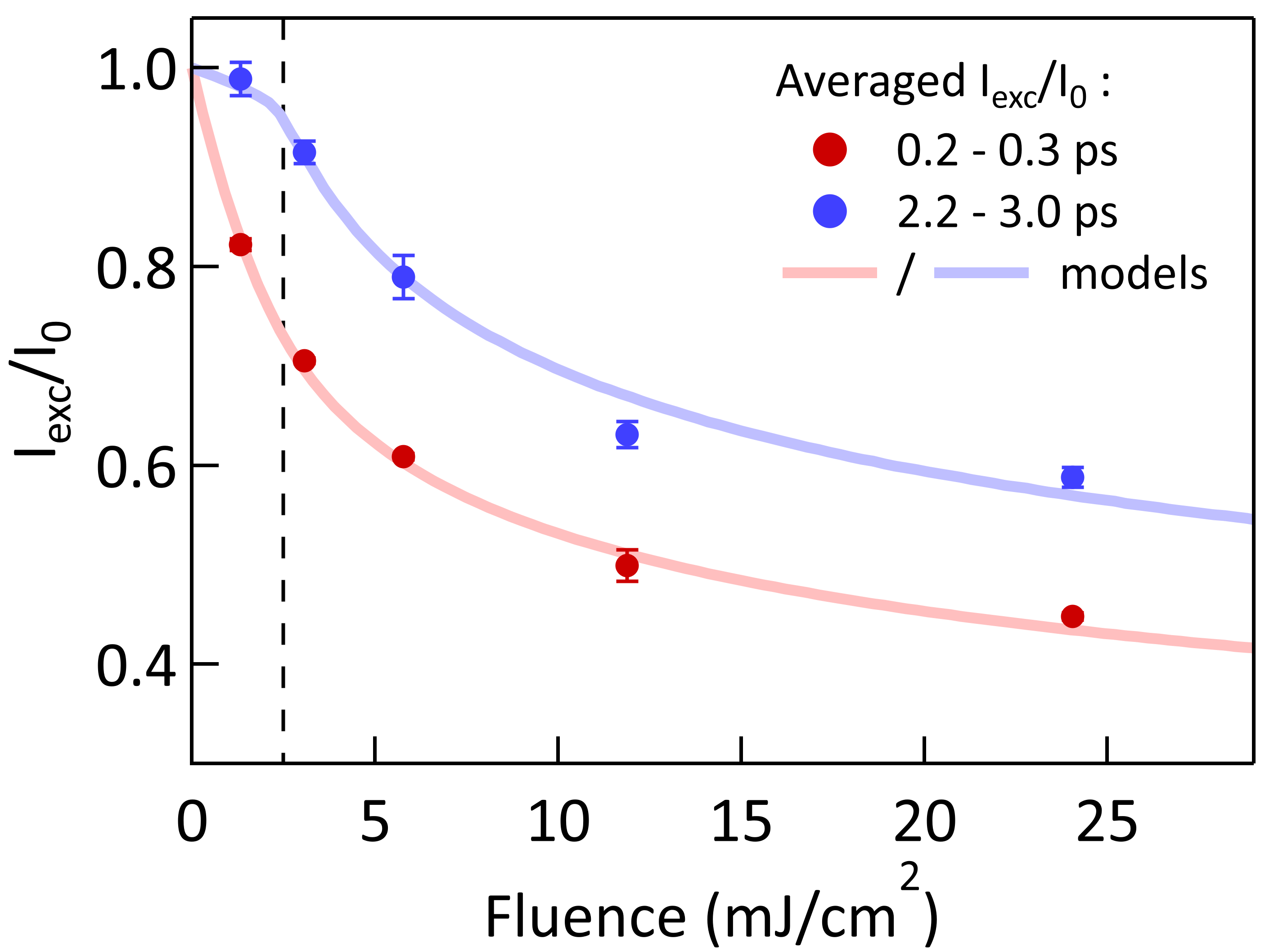}
\caption{Absorbed fluence dependence of the normalized intensity at 0.25~ps and at 2.65~ps. Error bars represent the standard deviation of the measurement points gathered in the vicinity of the nominal value for each fluence. The dashed line specifies the fluence corresponding to the critical energy density at the surface. Solid lines are obtained by using the model described in the text.}
\label{fig:peaks_orderparam}
\end{figure}

Finally, the model, taking into account the experimental resolution, enables us to extract the melting timescale of the \afm order, estimated in $\hat{\tau}=62(12)$~fs.
The resulting timescale of the melting of the \afm order following a fs laser excitation is faster compared to other 5$d$ oxide materials~\cite{Dean2016,AfanasievPRX9,Mazzonearxiv} (see Table~\ref{tab:table1}) and, in agreement with the estimate based on Eq.~(\ref{eq:erf_rec}),  and it is comparable with the experimental resolution.
 % see a discussion in PRB 86, 035128 (2012) for thermal and cooling of the free carriers, though here we use higher fluences...
The critical behaviour of the squared order parameter $\eta^2$ as a function of the excitation fluence is illustrated in Fig.~\ref{fig:peaks_orderparam}, which compares the measured relative changes of the scattering intensity at 0.2-0.3~ps and 2.2-3~ps after later laser excitation. 
\textcolor{black}{The fluence $F_c$=2.5(2) mJ/cm$^2$ corresponding  to the critical value of the energy density $n_c$ at the surface has a value similar to that observed in the manganites~\cite{Beaud2014,PorerPRB101} and for the melting of the charge ordering in NdNiO$_3$~\cite{EspositoSD2018}, which has also an \MIT~coinciding with \afm ordering. The photoinduced magnetic response in the nickelate shows as for NaOsO$_3$ a prompt recovery of the magnetic ordering for low laser fluences ($<$1mJ/cm$^2$), suggesting that the critical fluence for this nickelate is lower than for NaOsO$_3$. } However, one should exercise caution in comparing fluence values from different experiments as they strongly rely on values of laser power and beam size measurements that can be affected by significant uncertainties.

\section{Discussion}
There is a consistent body of evidence that impulsive laser excitations above the band gap are able to drive an \MIT \textcolor{black}{ \cite{Fiebig2000,OgasawaraPRB63,OgasawaraJPSJ71,RuelloPRB76}}~and, for samples which concomitantly order antiferromagnetically, destroy the \afm long range ordering~\cite{CavigliaPRB88,Dean2016,AfanasievPRX9,Mazzonearxiv,Stoicaarxiv}. 
In these materials, one expects to be able to use time-resolved experiments to draw conclusions on the hierarchy of interactions leading to the development of the \MIT. Specifically, it is interesting to see if the \MIT~in oxides based on transition metal atoms could be categorized following the observed changes in physical properties in response to impulsive laser excitation.
For example, NaOsO$_3$ originally attracted interest as it was thought to be a rare example of a Slater insulator~\cite{CalderPRL108,GurungPRB18}, namely showing a magnetically driven \MIT. While alternative explanations of the nature of the \MIT~have been proposed~\cite{KimLifshitzPRB94,vale2018crossover,vale2018evolution}, it is clear that there is a strong coupling between the magnetism and electronic structure.\textcolor{black}{ \cite{CalderPRL108, middey2014, GurungPRB18, KimLifshitzPRB94}.  } 

It would therefore be interesting to compare the sub-100~fs dynamics in NaOsO$_3$ with those of NdNiO$_3$ in which there is a large consensus on the fact that the appearance of the \afm phase is a byproduct of the concurrent \MIT. NdNiO$_3$ is a charge transfer insulator (a Mott insulator has a gap between two bands of the same character, e.g. both 3d, whereas for a charge transfer insulator the gap has a mixed character, e.g. between the oxygen 2p and the transition metal 3d bands) and one would expect the electronic response time $\tau{_e}$ to be faster than $\tau{_m}$, the magnetic one.
Vice versa, for NaOsO$_3$ one would expect $\tau{_m}$ to be comparable with $\tau{_e}$. In this respect, neither our data nor those presented in Ref.~\onlinecite{CavigliaPRB88,Stoicaarxiv} are conclusive as they are limited by time resolution available in the time-resolved X-ray measurements.
Our experiments on NaOsO$_3$ have shown that the magnetic response to a fs optical excitation occurs within few tenths of fs after the fs laser excitation. Modelling the time trace with Eq.~(\ref{eq:erf_rec}) gives 
$\tau{_m}\sim$71(6)~fs, while the model described in Sec.~\ref{ResultsX}, which takes into account the experimental time resolution gives $\tau{_m}\sim$62(12)~fs. These value are comparable with those of $\tau{_e}$ in NaOsO$_3$ and NdNiO$_3$, which are also limited by the time resolution of the optical experiments (see Table~\ref{tab:table1}). 
While from our measurements at room temperature it would be tempting to conclude that $\tau{_m} >\tau{_e}$, it must be noted that the magnetization time response $\tau{_m}$ in \afm Sr$_2$IrO$_4$ can be halved when experiments are conducted at low temperatures T$_m$ sufficiently far from T$_N$~\cite{AfanasievPRX9}. 

Regrettably, drawing conclusions on the microscopic parameters governing the timescale of the disappearance of the \afm order parameter upon impulsive laser excitation is difficult, due to the scarcity of experiments reporting true timescales of the \afm response. The fact that this timescale seems to be faster in NaOsO$_3$ than other Ir based oxides could possibly point to the role played by the presence of itinerant magnetism facilitating the electron mobility and therefore the further delocalization of the magnetic moments after the fs laser excitation. However, recent experiments on FeRh, a metallic \afm at room temperature, suggest that changes in the band structure occur only on the order of few hundreds of fs~\cite{Pressacco2021}. So it is clear that the full details of the band structure should be taken into account in order to obtain a quantitative description of the temporal evolution of the \afm order parameter.
In this respect, it must be also mentioned that non-equilibrium dynamic mean field theory predicts for $\tau{_m}$ timescales on the order of a few fs.~\cite{AfanasievPRX9}  
Accordingly, in order to gain a deeper understanding of the interplay between the electronic structure and magnetism in oxide materials upon laser excitation more experiments with better time resolution are required.

%\textcolor{red}{to be continued... THz small field response can be modelled with  Rothwarf–Taylor model? see Ref.~\onlinecite{ChuNM2017}}

\begin{table}[t]%The best place to locate the table environment is directly after its first reference in text
\caption{\label{tab:table1}%
This table illustrates the decay constants $\tau_{m}$  and $\tau_{e}$, determined by Eq.~(\ref{eq:erf_rec}), associated with the magnetic and electronic degree of freedom, respectively. Selected oxides compounds, whose timescale of the
melting of the \afm phase has been reported, are included. T$_m\,$(K) indicates the temperature at which the pump-probe experiments on the \afm order parameter were performed. The ratio T$_{m}$/T$_N$ is also reported, as it could affect the measured value of $\tau_{m}$.
}
\begin{ruledtabular}
\begin{tabular}{lccccc}
\textrm{Sample}&
\textrm{$\tau_{m}$\,$(fs)$ }&
%\multicolumn{1}{c}{\textrm{$\tau_{R}$}}&
\textrm{$\tau_{e}$\,$(fs)$ }&
\textrm{T$_m\,$(K) }&
\textrm{T$_{m}$/T$_N$ }&
\textrm{Reference}\\
\colrule
Sr$_3$Ir$_2$O$_7$ & $<$120\footnote{Limited by experimental time resolution.} & $<$95\footnotemark[1] & 110 & 0.39 & [\onlinecite{Mazzonearxiv}, \onlinecite{ChuNM2017}]\\
Sr$_2$IrO$_4$ & 330 & 250 & 75 & 0.39\footnote{For the thin film sample used in Ref.~[\onlinecite{AfanasievPRX9}] T$_N \sim$~195~K.} & [\onlinecite{AfanasievPRX9}] \\
Sr$_2$IrO$_4$ & 560 & 250 & 190 & 0.97\footnotemark[2] & [\onlinecite{AfanasievPRX9}] \\
NdNiO$_3$ & $<$125\footnotemark[1]  & $<$60\footnotemark[1] & 40 & 0.27 & [\onlinecite{Stoicaarxiv}, \onlinecite{LiangJPD_2018}]\\
NaOsO$_3$ & 71(6) & $<$90\footnotemark[1] & 300 & 0.73 & this work\\
\end{tabular}
\end{ruledtabular}
\end{table}

%Other compounds timescale with erf((t-t0)/$\tau$):
%$Sr_3Ir_2O_7$: $\tau$ = 112 fs, $\tau_{FWHM}$ = 187 fs
%$Sr_2IrO_4$: $\tau$ = 333~fs, $\tau_{FWHM}$ = 576~fs
%$NaOsO_3$: $\tau$ = 80~fs, $\tau_{FWHM}$ =  133~fs
%$NdNiO_3$: $\tau$ = 260~fs, $\tau_{FWHM}$ = 433~fs

\section{Conclusions}
In summary, we have presented the results of time-resolved X-ray experiments on NaOsO$_3$, which aimed to ascertain the time scales of the melting of the \afm structure. We have found that the magnetic order is quenched on a timescale $\tau{_m} \sim 71(6)$~fs, which is faster than in other perovskite compounds. These results may indicate the strong coupling between the electronic and magnetic degrees of freedom in this material. \textcolor{black}{ Our measurements of changes in the structural Bragg reflections show evidence of subsequent lattice dynamics extending over times of several picoseconds.}

The raw data files that support this study are available via the Zenodo repository~\cite{zenodo1}.

\section*{Acknowledgments}
We thank D. Hsieh and D. Mazzone for sharing their data, which were used to extract some of the timescales $\tau$ reported in Table~\ref{tab:table1}. We thank I. Lo Vecchio for sharing the NaOsO$_3$ data from Ref.~\onlinecite{LoVecchio2013}. We are also indebted with J. Mentink and M. Eckstein for stimulating discussion. This research work was supported from funding provided by the Swiss National Science Foundation, SNF Project No. 200021$\_$162863. Work in Japan was partly supported by JSPS KAKENHI Grant Number JP20H05276, and Innovative Science and Technology Initiative for Security (Grant Number JPJ004596) from Acquisition, Technology, and Logistics Agency (ATLA), Japan. The experiment at SACLA was performed with the approval of the Japan Synchrotron Radiation Research Institute (JASRI; Proposal No.~2017A8021). S.S. acknowledges ETH Zurich Post-Doctoral fellowship and Marie Curie actions for People COFUND program and the Carl Tryggers foundation with project number (113705531).

\section*{Appendix}

\subsection{Modelling of the antiferromagnetic X-ray diffracted intensity}
\label{sec:XDImodel} 
In a static diffraction experiment the measured intensity of a selected diffraction peak $I(q)$ is proportional to the square of the unit cell structure factor $F(q)$, with $q$ being the momentum transfer. If the sample is excited by an ultrashort laser pulse, the expression of the structure factor needs to be modified to take into account the depth-dependent excitation profile. Namely, close to the surface of the sample, the laser excitation would suppress or modify the scattering cross section, while layers further away from the surface could be unaffected, at least on very short time scales. It is customary to account for this depth dependence by splitting the sample into $N$ layers, each at a given depth z$_i$. Each layer will scatter X-rays differently and will contribute to the total diffracted intensity. To calculate the total diffracted intensity two limiting cases are typically considered. In the first case, the in-plane coherence length $\xi$ is much larger than the X-ray effective penetration depth $\zeta$ (the X-ray penetration depth $\zeta_X$ must be corrected by a geometrical factor that takes into account the scattering geometry, $\zeta \sim 3\zeta_X$ in our case). In this case the contributions from each layer are summed in amplitude and the diffracted intensity is given by
\begin{equation}
 I(q,t) \propto  \left| \sum_{i=0}^{N} F(q,t) \right|^2 .  
 \label{eq:S2}
\end{equation}
In the opposite case, where the penetration depth $\zeta$ is much larger than the in-plane coherence length $\xi$, the contributions of the different layers should be added incoherently such that the diffracted intensity is expressed as
\begin{equation}
 I(q,t) \propto  \sum_{i=0}^{N}  \left|F(q,t) \right|^2 . 
 \label{eq:S_F2}
\end{equation}
In our experiment we are in an intermediate situation, where $\xi$  and $\zeta$ are of the same order of magnitude, so it is not evident which would be the best approximation to describe the diffracted intensity. In order to model the experimental results, we have accordingly empirically fitted the data first with Eq.~(\ref{eq:S2}) and subsequently with Eq.~(\ref{eq:S_F2}). As shown by Fig.~\ref{fig:Maxmodel}, the model based on the expression of the intensity given in Eq.~(\ref{eq:S_F2}) reproduces well the fluence dependence of the antiferromagnetic diffracted intensity. Conversely, the model based on Eq.~(\ref{eq:S2}) does not provide a reasonable description of the measured diffraction intensities. We have therefore concluded that, for our specific experimental conditions, the model based on Eq.~(\ref{eq:S_F2}) is best suited to describe the data we have gathered on the antiferromagnetic peak.
\subsection{Lattice dynamics}
\begin{figure}[!tb]
%\centering
\includegraphics[width=\columnwidth]{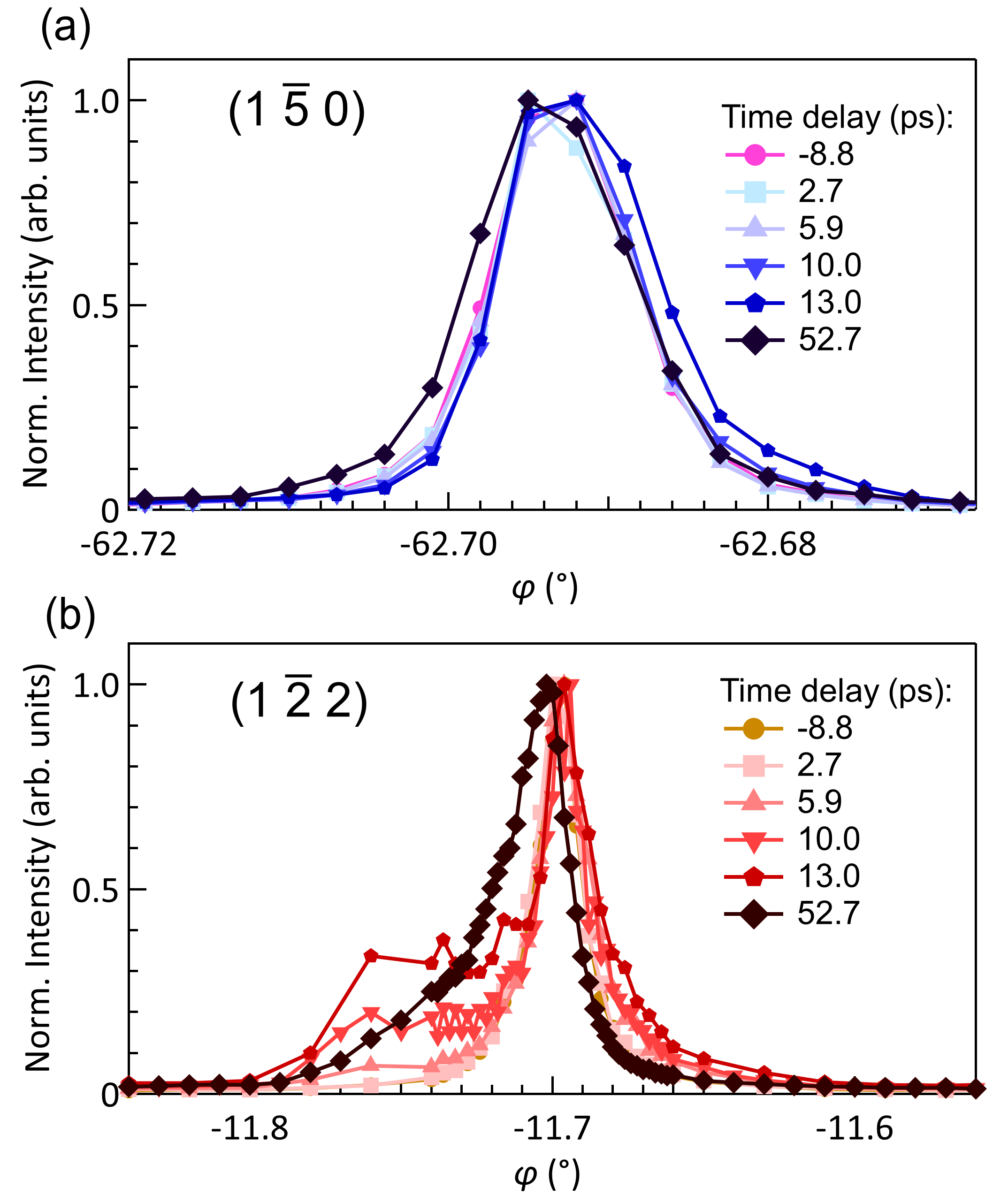}
\caption{Angular $\phi$-scans at different time delays at the maximum excitation fluence 24.2 mJ/cm$^2$ for the (1 $\bar{5}$ 0) magnetic and the structural (1 $\bar{2}$ 2) Bragg peaks. The intensity for each time delay has been normalized to maximum value within each scan for the ease of comparison of the change of the peak position and peak FWHM.}
\label{fig:phiscans}
\end{figure}

\begin{figure}[!tb]
%\centering
\includegraphics[width=\columnwidth]{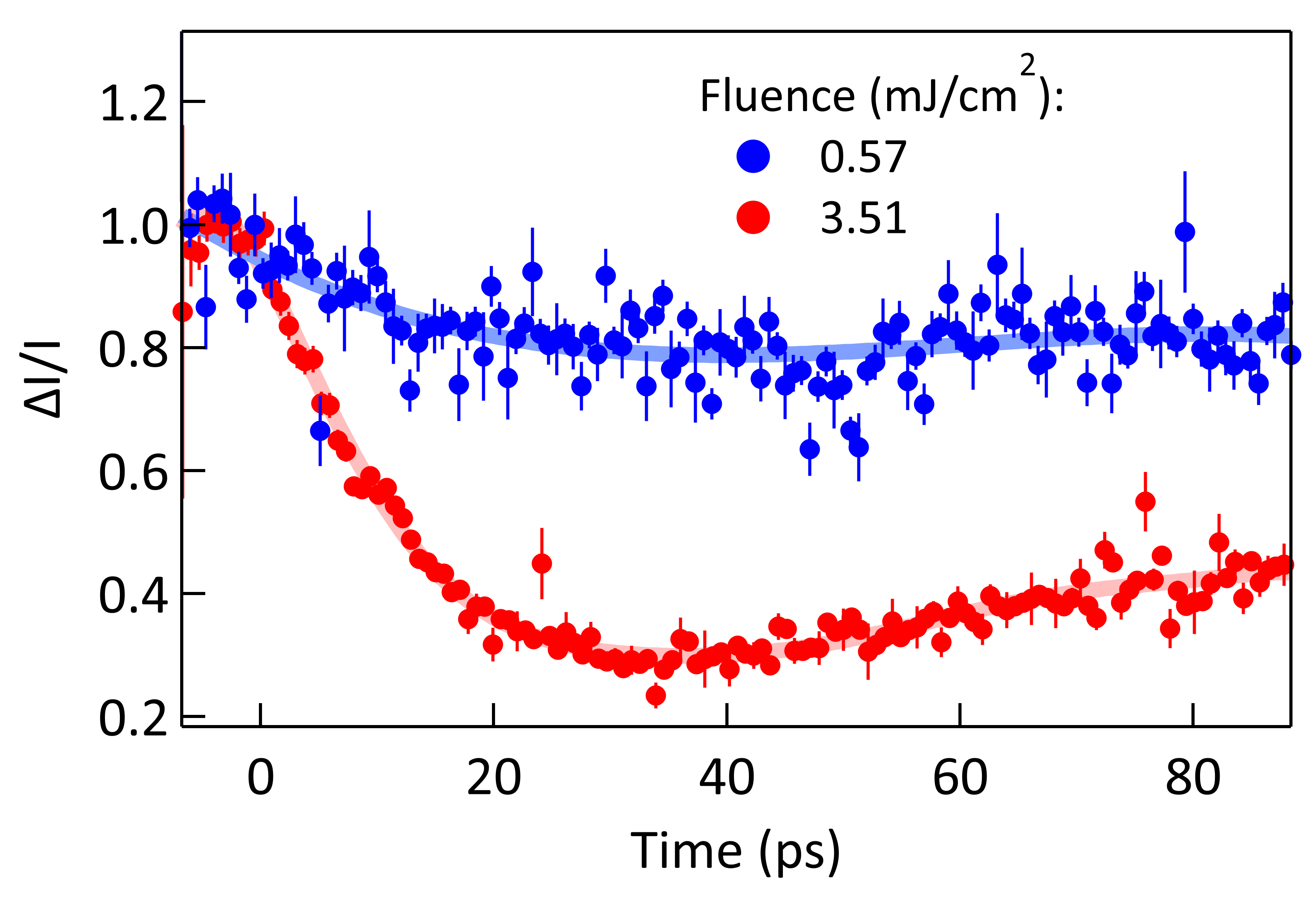}
\caption{Time traces of the change in intensity of the structural peak (2$\bar{4}$0) measured upon photoexcitation for two different fluences.}
\label{fig:peaks_lattice}
\end{figure}
To ascertain the presence of lattice deformation due to laser induced heating or strain waves we have performed angular scans at selected time delays on a magnetic and lattice peak. In order to maintain the X-ray grazing incident angle
fixed during the scan, we have scanned the diffractometer angle $\phi$. The results, gathered with the laser fluence of 24.05 mJ/cm$^{2}$ and normalized to 1 for the ease of comparison, are illustrated in Fig.~\ref{fig:phiscans}a for the magnetic peak (1 $\bar{5}$ 0) and in Fig.~\ref{fig:phiscans}b for the structural peak (1 $\bar{2}$ 2). 
The magnetic peak does not show an appreciable change of the peak position for time delays up to 10~ps. Also at time delay of several tenths of ps, where heat diffusion has occurred, the change in the peak position is tiny. Such a behaviour is not observed for the structural peak. As shown in Fig.~\ref{fig:phiscans}b, the position and the shape of the (1 $\bar{2}$ 2) Bragg peak changes significantly for time delays larger than 2.7~ps. Specifically, for time delay of $\sim$\,6~ps a second peak appears at $\phi=-11.76$, indicating the presence of a laser induced heating effect. At 10~ps time delay the presence of the second peak is more prominent and it is clearly visible as a distribution of intensity between the two peaks, which we ascribe to the presence of a thermal gradient in the sample. Finally at larger time delays, the heat has diffused from the lattice planes close to the sample surface to all of the X-ray probed volume, resulting in a sizable shift of the structural Bragg peak. A thermal gradient is still present, as indicated by the larger full width at half maximum (FWHM) and the asymmetric shape of the Bragg peak. 
The fact that the angular scans of the two peaks show such a marked difference for time delays larger than 2.7~ps can be understood as follows. The antiferromagnetic peak intensity decreases strongly as the sample temperature approaches T$_N$. Therefore, in the case of the magnetic peak no second peak nor a distribution of intensity is to be expected.  

As a next step, in order to determine the laser induced lattice dynamics, two structural lattice peaks (1~$\bar{2}~$2) (with no scattering contribution from the Os ion) and (2~$\bar{4}$~0) (with a scattering
contribution from the Os ion) were measured, using X-rays with an incident photon energy of 10.787~keV (0.115~nm), the same used for the \afm peak. Typically, phonon modes involving heavy ions are lower in frequency, and the laser excitation energy is transferred from electronic systems first to high energetic phonon modes. So one could expect a different time evolution for the two reflections, subsequent to the laser stimulus.
The timescales of the drop in lattice peak intensities were found to be $\tau_{\mathrm{O},\,\mathrm{Na}} \sim$11~ps (1~$\bar{2}$~2)
and $\tau_{\mathrm{Os}} \sim$ 20~ps (see Fig.~\ref{fig:peaks_lattice}), so significantly slower than the observed \afm order parameter and in a time window where coherent lattice expansion is expected to occur due to the laser heating effect in the excited sample volume. We tentatively ascribe the different dynamics observed to the different momentum transfer projection of the two reflections along the surface normal direction. Unfortunately, due to the limited amount of available measurement time, it was not possible to obtain more detailed information on the lattice deformation and a complete fluence dependence for such reflections. The observed dynamics is therefore ascribed to the expansion of the crystal lattice due to the heat deposited by the excitation laser pulses.

\ifx true false
\fi

\nocite{*}
%\bibliography{NaOsO3_UF}

\providecommand{\noopsort}[1]{}\providecommand{\singleletter}[1]{#1}%

\end{document}